\def\year{2021}\relax
\documentclass[letterpaper]{article} 
\usepackage{aaai21}  
\usepackage{times}  
\usepackage{helvet} 
\usepackage{courier}  
\usepackage[hyphens]{url}  
\usepackage{graphicx} 
\usepackage{natbib}  
\usepackage{caption} 
\usepackage[table,xcdraw]{xcolor}
\newcommand{\edit}[1]{\textcolor{black}{#1}}
\title{Images, Emotions, and Credibility: Effect of Emotional Facial Expressions on Perceptions of News Content Bias and Source Credibility in Social Media}

\author{
 Alireza Karduni\textsuperscript{\rm 1,\rm 2},  
    Ryan Wesslen\textsuperscript{\rm 2},
    Douglas Markant\textsuperscript{\rm 2},
    Wenwen Dou\textsuperscript{\rm 2}
    \\
}
\affiliations{
    \textsuperscript{\rm 1}Northwestern University\\
    \textsuperscript{\rm 2}University of North Carolina at Charlotte\\

}

\begin{document}

\maketitle

\begin{abstract}
Images are an indispensable part of the news we consume. Highly emotional images from mainstream and misinformation sources can greatly influence our trust in the news. We present two studies on the effects of emotional facial images on users' perception of bias in news content and the credibility of sources. In study 1, we investigate the impact of \edit{repeated} exposure to content with images containing positive or negative facial expressions on users’ judgments of source credibility and bias. In study 2, we focus on sources' continuous emotional portrayal of specific politicians. Our results show that the presence of negative (angry) facial emotions can lead to perceptions of higher bias in content. We also find that consistent negative portrayals of different politicians leads to lower perceptions of source credibility. These results highlight how implicit visual propositions manifested by emotions in facial expressions may have a substantial effect on our trust in news.
\end{abstract}

\section{Introduction}
In the aftermath of the US 2020 presidential election, some rioters who did not trust the official results of the election stormed the capitol building. Around the same time, when the first Covid-19 vaccines were introduced, many did not trust the news around the safety of the vaccines. Observing such incidents, what causes news consumers to not trust certain mainstream news sources while believing misinformation published by others? Recent research suggests one answer may be cognitive factors, such as the propensity for analytical reasoning affecting belief in the accuracy of news content \cite{pennycook2021psychology}. However, our judgments of news source credibility is also an important but often neglected factor shaping our judgments of \edit{news and misinformation} \cite{pehlivanoglu2021role, wallace2020sources}. Indeed, prominent misinformation scholars advocate for increasing focus on behaviors of news sources rather than individual stories since ``the defining element of fake news is the intents and processes of a publisher'' \cite{lazer2018science}. In this research, we tackle the interactions between news sources and consumers by exploring how social media users' evaluations of the news content influence their judgments of news source credibility. \edit{We posit that identifying factors that influence consumers' judgments of news sources is an essential task, not only to help detect and combat misinformation, but also to identify and combat practices that reduce consumers' trust in more trustworthy 
news sources.} 

News sources often frame their content with specific styles, tones, and emotions. Both mainstream news media and misinformation sources take advantage of highly emotionalized and sensational content to influence their audiences \cite{arsenault2006conquering,wardle2017information,volkova2017separating,richards2007emotional, peng2018same,rathje2021out}. Various studies have shown that such content strategies are often effective. People are more likely to be drawn to highly emotionalized news, click on articles associated with extreme emotions, and share headlines with more negative sentiments \cite{bowman202017,reis2015breaking,berger2012makes}. In addition to emotional verbal content, news sources take advantage of social media’s highly visual nature by choosing images that amplify the persuasive power of the verbal content \cite{wardle2017information, grabe2009image}. Examples of such visual content include images that communicate racist concepts not present within the text \cite{messaris2001role}, emotional facial expressions of news-casters \cite{mullen1986newcasters}, \edit{and partisan media producing ideological bias in their visual coverage of political candidates with features including facial expressions and face size \cite{peng2018same}}.  

In this work, we tackle a salient aspect of visual information, facial expressions, in social media posts on users' judgments: Do users perceive a source as less credible when the source repeatedly publishes tweets with angry facial expressions? What if a source systematically portrays a negative visual bias against a politician (e.g. Bernie Sanders or Donald Trump) by repeatedly publishing news with angry images of those politicians? We explore these questions by considering \textbf{source credibility}, a primary factor in the persuasiveness of news sources \cite{hovland1953communication,wallace2020sources}; and \textbf{content bias}, defined as ``having a perspective that is skewed,'' \cite{wallace2020sources} as two primary elements of users' judgments of news sources. We investigate how repeated usage of images with positive (happy) or negative (angry) emotional facial expressions in social media posts affects users’ judgments about \textbf{source credibility} and \textbf{content bias}. 

We seek to answer the following questions: 1) How does repeated exposure to news content accompanied by images with positive (happy) or negative (angry) facial expressions impact users’ judgments about a source's credibility?; and 2) How sources' \edit{repeated} positive or negative visual portrayal of different politicians interacts with users' prior attitudes (favorability and familiarity) towards those politicians to shape their judgments of source credibility. 
To answer these questions, we conducted two consecutive preregistered controlled experiments:

\begin{enumerate}
    \item In study 1 we examined how users' credibility judgments are impacted by news sources that show a pattern of using highly emotionalized visual images.\footnote{Pre-registration: \url{https://aspredicted.org/blind.php?x=9rn6i7}} More specifically, we investigated how angry or happy facial expressions in tweet images affect users’ judgments about eight anonymized right/left-leaning and mainstream/misinformation news accounts.
    
    \item Leveraging findings from study 1, in study 2 we focused on  sources with repeated emotionalized visual coverage of influential politicians.\footnote{Pre-registration: \url{https://aspredicted.org/blind.php?x=9js6d5}} We examined how users' judgments are impacted by sources that consistently portray specific politicians' emotions as negative or positive. In study 2, we also investigated how users’ prior attitudes towards each politician interact with sources' negative or positive emotion portrayal of those politicians. 
\end{enumerate}

 In study 1, we observe that users find tweets with negative (angry) facial expressions as more biased. However, we found that \edit{negative facial expressions might not} lead to lower perceived source credibility and \edit{that this effect is likely heavily moderated by textual content and tone of the news coverage (e.g., misinformation sources were deemed as less credible)}. In study 2, we found that  \edit{focusing on mainstream} sources' repeated negative visual portrayal of specific politicians, users are more likely to perceive the content as more biased and the source as less credible. Across both studies, in addition to quantitative analysis of the experimental treatments, we qualitatively analyzed users' comments explaining reasons behind their judgments. Our qualitative analysis demonstrated that users also relied on several cues such as opinionated language, political bias, and satire to guide their credibility judgments. These studies highlight the extent to which highly emotionalized visual coverage might impact users’ trust in news sources.

\section{Related work}

\textbf{Source Credibility and Content Bias}: False, inaccurate or ``fake'' news is a serious problem for our societies and democracies \cite{lazer2018science,karduni2019vulnerable,wardle2017information}. In their work on decision-making about misinformation, Pennycook and Rand  highlight different aspects of why users fall for ``fake news.'' They provide evidence that ``contradict the common narrative that partisanship and politically motivated reasoning explain why people fall for fake news'' (\cite{pennycook2021psychology} pg. 1) Furthermore, they find that the propensity to engage in reasoning is a more important factor in users' ability to detect false news. 

In addition to users' ability to detect instances of inaccurate and false information, perceptions of source credibility influence how users interact with and are persuaded by the news \cite{pornpitakpan2004persuasiveness}. Perceptions of trustworthiness and expertise are noted as two main factors in credibility judgments \cite{petty1997attitudes,wallace2020sources}. Research on credibility perceptions on social media has highlighted that when faced with short content (e.g., Twitter), users might have different credibility perceptions for different topics (e.g. science versus entertainment) and might rely on different heuristics such as username style to judge source credibility \cite{morris2012tweeting}. Recent work by Laura Wallace suggests that perceptions of source bias and untrustworthiness can have different effects on perceptions of source credibility. For instance, in certain cases users' might believe sources are biased but honest in their reporting and therefore credible \cite{wallace2020sources,wallace2020influences}. 

\textbf{Images, Emotions, and Users' Judgments of News}: Visual information impacts how users judge the credibility of information sources in several ways. Wobbrock et al. show that visual appearances like font size or inclusion of images may impact users' perceptions of credibility \cite{wobbrock2021goldilocks}. Different images may have varying effects on users' judgments. For example, after seeing an image of a brain, users are more likely to believe claims in certain scientific articles \cite{schweitzer2013fooled,mccabe2008seeing} while images of smoking increases belief in messages in warning signals \cite{shi2017importance}. Articles accompanied by alarming images \cite{knobloch2003imagery} or ones that depict victimization \cite{zillmann2001effects} increase users’ selective interaction with the articles. Furthermore, when exposed to false versus true news headlines, participants' self-reported experience of heightened emotion (users who report feeling more angry or sad) is linked to an increase in perceived accuracy of false news but not factually correct news \cite{martel2020reliance}.

Recent work in social psychology \edit{and media bias} has highlighted the interpersonal effect of emotions on users' decisions \cite{van2010interpersonal}. \edit{For example, displays of angry (but not happy) facial expressions, might reduce the likelihood of our decisions to trust others \cite{campellone2013you}}. \edit{Moreover, displays of different emotions can impact perceptions of cooperativeness and appraisal (e.g., happy agent perceived to be more likely to cooperate and angry/sad the opposite) \cite{de2012reverse}}. Andalibi and Buss assert that people find emotions as ``insights to behavior, prone to manipulation, intimate, vulnerable, and complex'' \cite{andalibi2020human}. Emotional content in images also impacts the likelihood of people believing a statement. For example, being exposed to highly negative emotional images about different phenomena increases belief in the accuracy of news content \cite{vlasceanu2020emotion}. In comparison to positive imagery, being exposed to negative imagery has been associated with greater perceived accuracy of false information \cite{porter2010prospective}. Melodramatic animations are also shown to likely increase users' perceptions of credibility of a news report \cite{lo2017use}. \edit{From the perspective of visual media bias, \cite{grabe2009image} called for more attention on studying the visual framing of politicians in news due to the increase in visual (image) portrayals in news. A study by \cite{peng2018same} showed that, among other features, happy facial images increase the perceived favorability of a political candidate and perceived trustworthiness, but angry facial expressions decrease both.} Another study by Masch and Gabriel suggests that depending on the political context, positive and negative depictions of politicians in videos might impact users' attitudes towards those politicians \cite{masch2020emotional}.

Current research on misinformation focuses on perceptions of content accuracy and users' ability to distinguish false and true information \cite{pennycook2021psychology} based on various cues including emotions in text \cite{karduni2018can} or images \cite{vlasceanu2020emotion}. Research on persuasion has shown users' judgments of source credibility and content bias, and the certainty of those judgments, are important factors in news sources' persuasive power \cite{wallace2020sources,tormala2004source}. Inspired by the mentioned work, this paper studies an important but, to our knowledge, overlooked aspect of our \edit{judgments of news sources}: the effect of sources' \edit{repeated and potentially} systematic usage of highly emotional facial images on users' judgments of source credibility and content bias.

\section{Study design and implementation}
 Considering perception of source credibility as a product of the news content that users consume, we formulate the overall structure of the studies\footnote{Approved by our institutions' Internal Review Board (IRB)} as 1) Users first view curated social media posts from an \textbf{anonymized} news source; 2) Users evaluate bias for multiple posts from the source; 3) After deciding they have viewed and evaluated enough posts from a source (with a hard minimum of five tweets), users are instructed to provide a credibility rating of the source. Users repeated these steps for 8 different new sources.
 
 This structure applies to both studies 1 and 2. The randomized treatments were applied by altering the content users see from each source (i.e. content with positive or negative images, or no images as a control condition).
 
As opposed to limiting users judgments to binary outcomes, we elicited users' bias and credibility judgments on continuous response scales. We elicited perceived bias of a tweet as a continuous number between 0 (unbiased) and 1 (biased). Similarly, we elicited perceived credibility of a source as a number between 0 (not credible) and 1 (credible). As it has been argued that user attitudes with higher certainties have important consequences including increasing resistance to persuasion and being persistent over time \cite{tormala2004source,tormala2007attitude}, 
we elicited users’ certainty around their judgments as a confidence interval within each respective domain. We adopted a modified version of the Line + Cone technique \cite{karduni2020bayesian} to elicit users' judgments. The visual technique (called Line + Range; See Figure \ref{fig:rq2-sample}) enabled us to elicit judgments and uncertainty using a visual technique. To explore users' self-described reasons behind their judgments, we also asked users to provide verbal comments using an open text box.



\subsection{Dataset and Study Stimuli}
The tweets dataset was collected from the Twitter streaming API from October 25th to January 25th, 2018 from multiple news accounts labeled as misinformation or mainstream by Karduni et al. \cite{karduni2019vulnerable,karduni2018can}. After downloading all images from the dataset, we used the python face-recognition library\footnote{\url{https://pypi.org/project/face-recognition/}} to extract images with faces and cropped the images to only include faces. To identify images of different politicians, we used Google’s FECNet \cite{schroff2015facenet} to extract feature vectors of cropped faces and used HDBScan \cite{mcinnes2017hdbscan} to cluster faces and manually identified influential politicians. \edit{We chose to include only cropped faces in our study to remove potential confounding factors such as body language and surrounding scenes, as studies show that recognition of facial expressions is influenced by both
\cite{kret2013.00810, Kret2012a, Kret2013}.}

To extract emotion predictions from faces, we used two separate python libraries, EmoPy \footnote{https://github.com/thoughtworksarts/EmoPy} and deepface\footnote{https://pypi.org/project/deepface/}, which provide the ability to sort the tweets based on emotion predictions and model confidence. After generating a candidate dataset, we qualitatively excluded images that were low-resolution, inaccurately assigned to a cluster, or had wrong emotion predictions, 
All qualitative coding to produce the final datasets for both studies were done by the authors. The first author first created a larger candidate dataset. Using the study interface, the other authors iteratively evaluated each data point. Disagreements were resolved in team meetings. 

\subsection{Study 1 \& 2 models}
 Our experimental design for both studies consisted of repeated measures for each user within two levels of responses including source level responses (credibility) and tweet level responses (bias). Users' responses were continuous variables bounded between 0 and 1. For both studies, we used mixed-effects beta regressions with \texttt{glmmTMB} to address the hierarchical design of our studies and the bounded dependent variables. Within the text, model coefficients are reported as log-odds ratios with corresponding confidence intervals. In the model figures, we transformed the log-odds ratios to odds ratios for ease of interpretation. Odds ratios show the direction and strength of how each independent variable impacts the dependent variables. We used the normal approximation to calculate p-values of fixed effects and t-values produced by \texttt{lme4}.



\section{Study 1}

\begin{figure}[t]
  \centering
    \includegraphics[width=0.6\columnwidth]{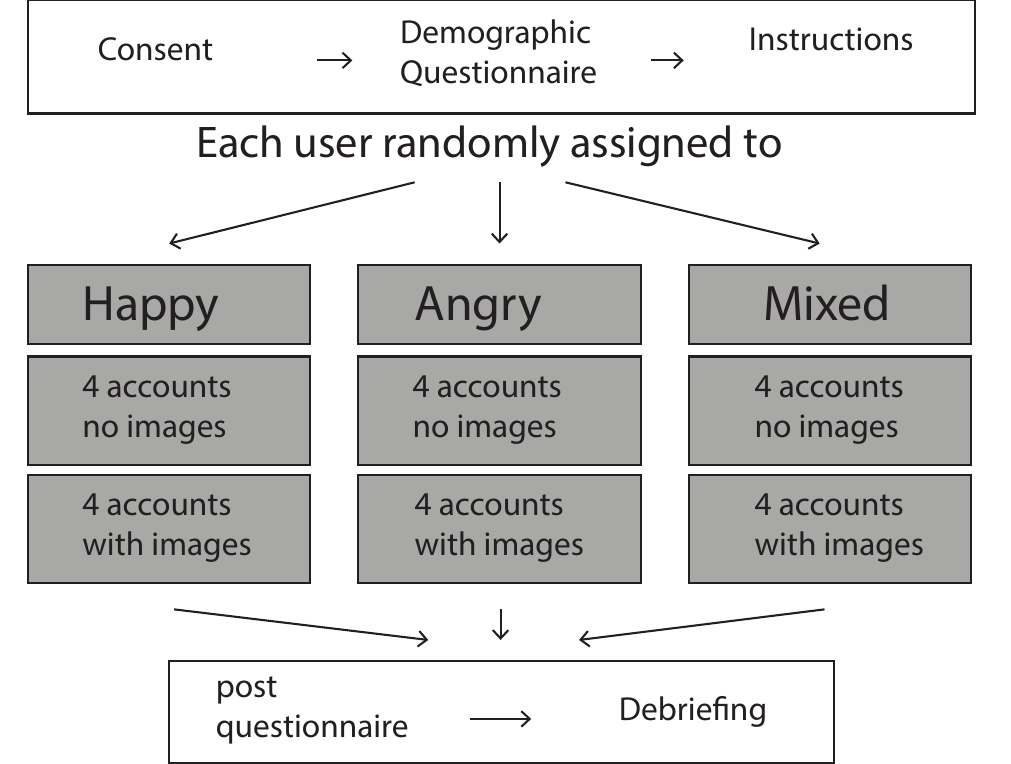}
  \caption{Study 1 conditions and process
  }
  \label{fig:rq1-design}
\end{figure}

In Study 1 we examined how positive (happy) and negative (angry) emotions in images influence perceived content bias (tweets) and source credibility (accounts).
Participants completed a task in which they observed a series of tweets from eight different accounts. To provide a variety of different political orientations and content types,
the accounts were either categorized as \textbf{mainstream} or known to produce \textbf{misinformation}. 
The accounts were also categorized as being right-leaning or left-leaning (see Table \ref{study1Accs}).\footnote{Right or left-leaning labels were confirmed with \url{https://mediabiasfactcheck.com/} and \url{https://www.allsides.com/}} Given these two dimensions of mainstream/misinformation and right/left political orientation, users evaluate a total of eight accounts with more than 90 tweets in the final dataset. 
 The content shown to users was manipulated in a between-subjects manner. Users were randomly assigned to one of three conditions: \textbf{happy}, \textbf{angry}, and \textbf{mixed}  (see Figure \ref{fig:rq1-design}). The happy condition contained tweets with the highest happiness scores for the images. The angry condition included tweets containing images with the highest angry image scores. The mixed (control) condition alternates between happy and angry images. Furthermore, one account from each category (right, left, misinformation, mainstream) was randomly selected to include no images. 
For example, a participant assigned to a happy condition,  viewed tweets with the highest rated happy images from four of the eight accounts but did not see images from the other four accounts.  

For each account, users responded based on their judgments about the bias of each tweet. By clicking on ``View more tweets'', participants viewed additional tweets until they decided to rate the credibility of the source (a minimum of 5 tweets was enforced for each account). 
By clicking on ``Make a decision,'' users saw a pop-up view in which they entered their credibility judgments. Users also had the option to use a text box to describe what influenced their decisions.

\begin{table}[t]
\scriptsize	
\sffamily
\centering
\caption{8 anonymized Twitter accounts in Study 1}
\label{study1Accs}
\begin{tabular}{|l|l|l|}
\hline
\rowcolor[HTML]{656565} 
{\color[HTML]{FFFFFF} Source name}         & {\color[HTML]{FFFFFF} Type}          & {\color[HTML]{FFFFFF} Orientation} \\ \hline
{\color[HTML]{000000} @veteranstoday}   & {\color[HTML]{000000} misinformation} & {\color[HTML]{000000} left}                  \\ \hline
\rowcolor[HTML]{C0C0C0} 
{\color[HTML]{000000} @amlookout}       & {\color[HTML]{000000} misinformation} & {\color[HTML]{000000} right}                 \\ \hline
{\color[HTML]{000000} @opednews}    & {\color[HTML]{000000} misinformation} & {\color[HTML]{000000} left}  \\ \hline
\rowcolor[HTML]{C0C0C0} 
{\color[HTML]{000000} @InvestWatchBlog} & {\color[HTML]{000000} misinformation} & {\color[HTML]{000000} right}                 \\ \hline
{\color[HTML]{000000} @MotherJones} & {\color[HTML]{000000} mainstream}     & {\color[HTML]{000000} left}  \\ \hline
\rowcolor[HTML]{C0C0C0} 
{\color[HTML]{000000} @Jeresulem\_Post} & {\color[HTML]{000000} mainstream}     & {\color[HTML]{000000} right}
 \\ \hline
{\color[HTML]{000000} @cnnPolitics} & {\color[HTML]{000000} mainstream}     & {\color[HTML]{000000} left}  \\ \hline
\rowcolor[HTML]{C0C0C0} 
 {\color[HTML]{000000} @nypost}      & {\color[HTML]{000000} mainstream}     & {\color[HTML]{000000} right}                \\ \hline
\end{tabular}
\end{table}

\subsection{Hypotheses} We hypothesize that in comparison to tweets with happy images, users are more likely to assess tweets with angry images as biased.
Furthermore, we also hypothesize that as compared to the mixed condition, users in the angry condition are more likely to assess sources as less credible. 
 However, since it is likely for users’ judgment to be influenced by the inherent differences in the nature of sources, we also explore the effects of political orientation (right vs. left) and source type (mainstream vs. misinformation) on their judgments. 

\subsection{Dependent \& Independent Variables} We considered four total dependent variables (DV): (1) content bias (bounded value between [0,1]), (2) uncertainty range around content bias (bounded value between [0,1]), (3) source credibility (bounded value between [0,1]), (4) uncertainty around source credibility (bounded value between [0,1]). For our independent variables (IV), we included the image emotion condition (angry, happy, or mixed), image shown (true or false), as well as political orientation (right or left), and source type (mainstream or misinformation). For models built based on tweet-level responses, i.e. bias choice/uncertainty of tweets as the dependent variable, only two image emotion conditions of happy or angry are applicable since each tweet contains at most one type of emotion. 

\subsection{Model Specification}
For each model, we included users’ unique id and the source id as random effects. The reference conditions for credibility choice and uncertainty models are image emotion = mixed, source orientation = left, source type = mainstream, and image shown = False. The reference conditions for bias choice and uncertainty models are image emotion = happy, source orientation = left, source type = mainstream, and image shown = False. 

\begin{figure}[b]
  \centering
    \includegraphics[width=0.9\columnwidth]{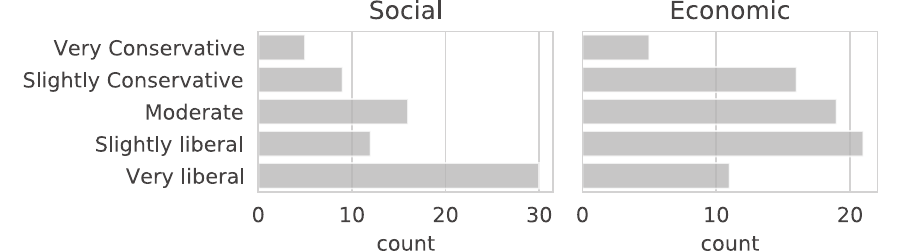}
  \caption{Participants' responses to questions ``How would you describe your political outlook with regard to economic / social issues?" for Study 1.
  }
  \label{fig:rq1-political}
\end{figure}

\subsection{Participants}

In this study's pre-registration, we originally planned to recruit up to 300 participants from crowdsourcing platforms. However, due to concerns about the quality of data collected from MTurk and time constraints, we shifted our participant pool to university students. We recruited a total of 81 university students with an average age of 21 years old. Per our pre-registration, we excluded responses from 9 participants with missing responses (due to unexpected technical difficulties), resulting in 72 accepted responses (48 woman, 23 man, and 1 other; 40 white, 13 African American, 7 other Asian, 4 East Asian, 5 Hispanic, 2 middle eastern, and 1 native American). \edit{Participants were highly liberal in regards to social political issues and more balanced in regards to economic issues (see Fig \ref{fig:rq1-political})}. Per random assignment, 30 participants were assigned to the angry condition, 24 participants to the happy condition, and the remaining 18 to the mixed condition. Participants took an average of 26 minutes to complete the study. All participants either received course extra credits or required research credits as incentives.

\begin{figure}[t]
  \centering
    \includegraphics[width=0.9\columnwidth]{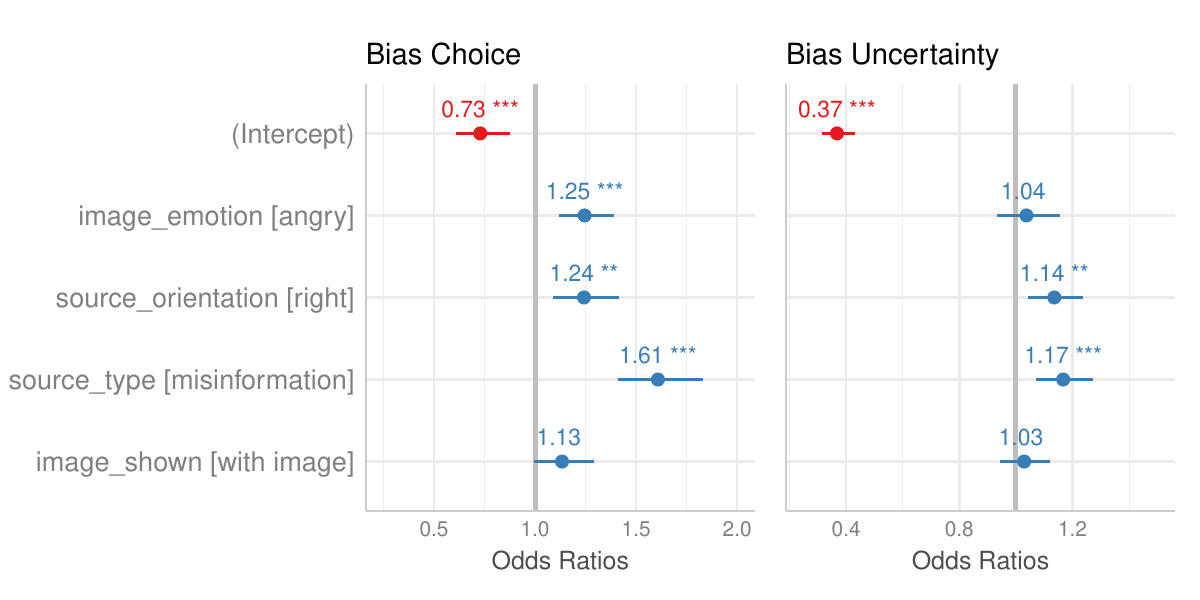}
  \caption{
Study 1 fixed effects odds ratios for bias choice (left) and bias uncertainty (right). Error bars indicate 95\% confidence intervals. 
Asterisks indicate statistical significance than zero using p-values: *** 99.9\%, ** 99\%, * 95\%. 
  }
  \label{fig:rq1-bias}
\end{figure}

\subsection{Results}

\subsubsection{Content bias:}

We used two mixed-effects beta regressions to study effects on users’ judgments about the bias of individual tweets (see Fig \ref{fig:rq1-bias}-left). Users found content from right-leaning sources to be more biased ($\beta = 0.216 \; [0.086,  0.34]$, $z = 3.25 , p < .01$). Similarly, users judged content from misinformation sources to be more biased ($\beta = 0.474 \; [0.34,  0.60]$, $z = 7.11, p < .001$). We also found that users rated tweets containing angry images as more biased ($\beta = 0.219 \; [0.11, 0.32]$, $z =  3.94, p < .001$). This is in line with our hypothesis that angry imagery will lead to an increase in users’ perceived bias. Surprisingly, even though the content shown to users was solely sorted based on facial emotions in images, we did not find a noticeable effect of emotional images on users bias ratings. It is possible that sources choose images to amplify tweets' content and users' judgement were not based on the emotion in facial expressions alone. 

In our mixed-effects model on users' uncertainty around their judgments (see Fig \ref{fig:rq1-bias}-right), we found that compared to left-leaning mainstream accounts, users were more likely to have larger uncertainty ranges for right-leaning accounts $(\beta = 0.127 \; [0.041,  0.21]$, $z=2.90  \; p < .01)$. We also found a similar positive effect effect for misinformation accounts ($\beta = 0.154 \; [0.06,  0.24], z = 3.52, p < .001$). 

\begin{figure}[t]
  \centering
    \includegraphics[width=0.9\columnwidth]{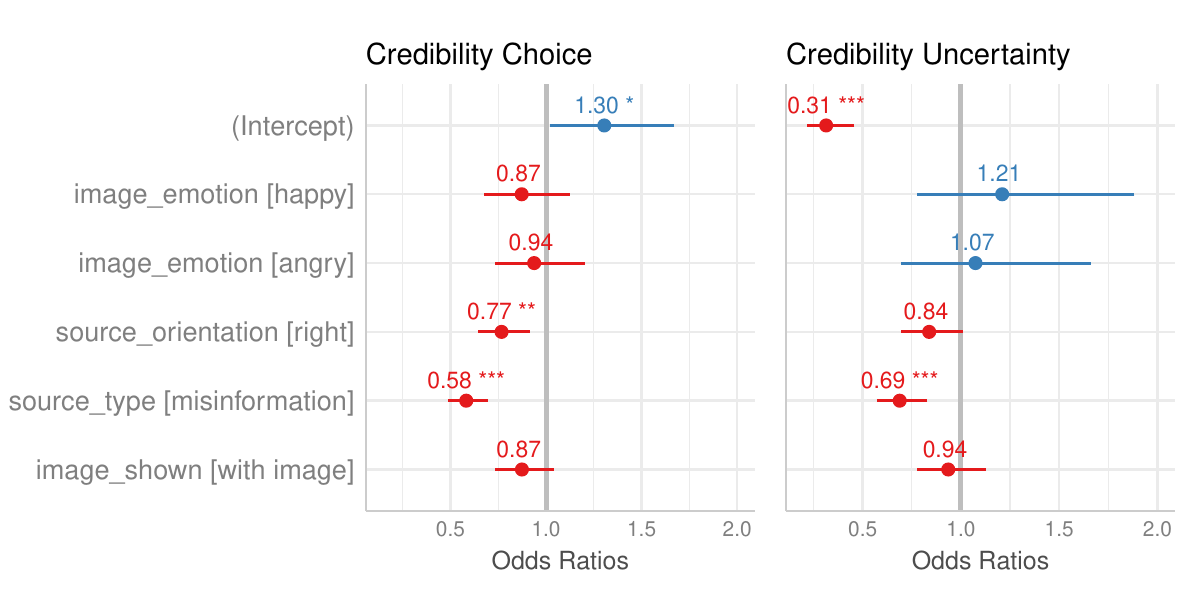}
  \caption{
Study 1 fixed effects coefficients for credibility choice (left) and credibility uncertainty (right). Error bars indicate 95\% confidence intervals. 
Asterisks indicate statistical significance than zero using p-values: *** 99.9\%, ** 99\%, * 95\%. 
  }
  \label{fig:rq1-credibility}
\end{figure}

\subsubsection{Source credibility:}
We used mixed-effects beta regression to investigate the effects of study conditions on users’ judgments about source credibility (see Fig \ref{fig:rq1-credibility}-left). We found that in reference to left-leaning accounts, users were more likely to rate right-leaning accounts as less credible ($\beta = -0.264 \; [-0.43, -0.08], z =  -2.961, p < .01$). Users were also more likely to rate misinformation sources as less credible ($\beta = -0.539 \; [-0.71, -0.36], z =  -6.024, p < .001$). We did not observe significant effects for the happy or angry conditions. In other words, we did not find evidence towards our hypotheses that being exposed to a source that repeatedly publishes tweets with angry or happy facial expressions impacts perceptions of source credibility.

We also used a mixed-effects beta regression to study whether users’ confidence around their decisions (represented through uncertainty ranges in the Line + Range technique). Interestingly, source type was the only factor that significantly affected users’ uncertainty on source credibility. Users were likely to be more confident in their decisions when source type was categorized as misinformation ($\beta = -0.373 \; [-0.55, -0.18], z = -3.921, p < .001$).

\subsection{Analysis of users' comments}

Each user had the option to answer one open-ended question for each account asking ``please describe how the tweets (text and images) influenced your decisions about this account?'' Even though leaving comments was an optional part of the study, we received comments from all 72 participants and the majority of participants left comments for all 8 trials. In total, we collected 572 comments about users’ decision-making influences. Since thematic analysis with a large number of comments was challenging, we used topic-modeling to facilitate the qualitative analysis of the comments and arrive at different themes of influences on users’ judgments. We categorized the extracted topics into four general themes: 1) sources' political orientation (left-leaning or right-leaning), 2) opinionated versus factual reporting, 3) specific language usage or tone (e.g. source is angry, or uses tabloid like language), and 4) the effect of images. Although the majority of users' comments were related to different cues tied to the text of the tweets, in this section we provide a summary of comments related to images.

Topics 9 (25 comments from 14 users) and 5 (19 comments from 10 users) included comments that mentioned images as a factor in their decision. These comments made a wide range of image-related observations from unflattering imagery, to facial expressions and lack of seriousness. For example, a user described how they perceived the image and the text to be not aligned: \textit{``Some of the tweets seemed to be about data rather than opinions. Some of the pictures seemed to take away merit.''} Another comment explicitly mentioned facial expressions of individuals helping them decide that the tweets have a left bias: \textit{``Many of the tweets seemed to be credible as most were quotes by others. Some of the pictures had facial expressions that made the tweets seem left swinging.''}

A number of comments cited comic or not serious imagery as the basis for their decisions. For example \textit{``Difficult to take the meme-like images seriously.''} And, \textit{``The images were cartoonish and difficult to take it seriously. It was obviously making fun of trump.''}. A few comments mentioned unprofessional or unflattering images as the basis for their judgments: \textit{``[...] and somewhat unflattering pictures of those who either commented against them or who may do (or not do) something against them.''} And, \textit{``Some of the pictures were not professional and showed the president and alliances in a negative light.''}

These comments show that at least for a group of users, images can serve as a cue for a source's lack of credibility. However, the majority of comments about pictures and images did not specifically mention emotions in images as the basis for their decisions. This might be due to the fact that tweets in our dataset contained a diverse set of topics. It is possible that users look for a more systematic negative treatment of specific topics or individuals, rather than a combination of negative imagery from a wide range of topics.

\begin{figure}[t]
  \centering
    \includegraphics[width=1.0\columnwidth]{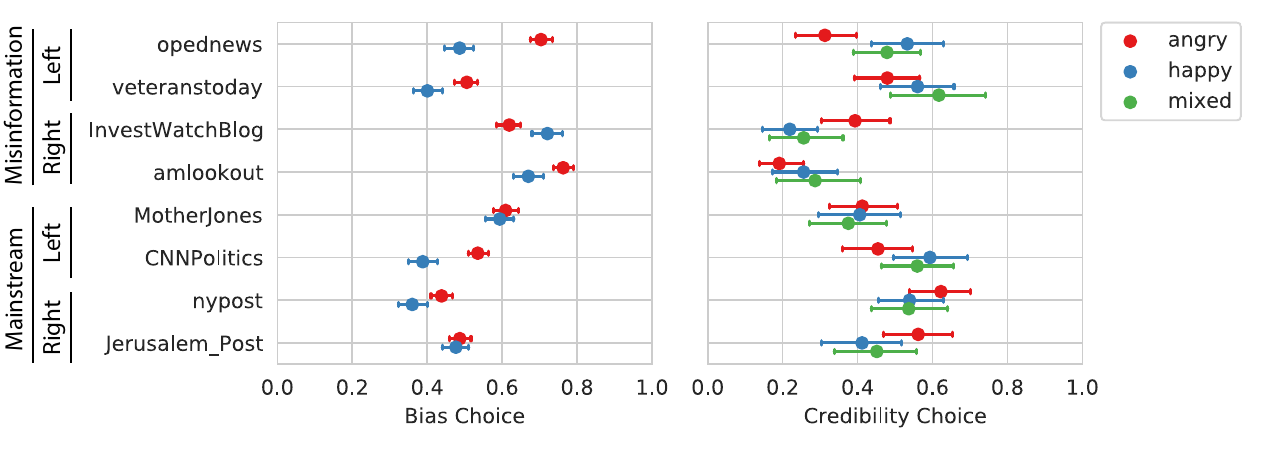}
  \caption{Mean and 95\% bootstrapped confidence interval for bias and credibility choice for different news accounts (sources) in study 1.
  }
  \label{fig:rq1-pointplot}
\end{figure}

\subsection{Study 1 Discussion}

We hypothesized that sources' systematic posting of tweets with angry images would lead to higher perceived content bias and lower perceived source credibility. We also hypothesized that happy images would lead to a reverse effect. We found partial evidence for our hypotheses: we observed an increase in perceived bias for tweets containing angry images (see Fig \ref{fig:rq1-pointplot} for raw point estimates for each account). We did not find evidence that this perceived content bias led to lower judgments of source credibility. We believe this is in line with recent findings by Wallace and colleague asserting that higher perceived bias does not necessarily lead to lower perceived credibility of sources \cite{wallace2020sources}. \edit{However, perceptions of content bias were likely affected by the general tone and topical focus of individual accounts (note that $source\_type = misinformation$ had the largest effect on perceived bias.} 
Our qualitative analysis of comments also hints that users rely heavily on text content to judge both bias in content and credibility of sources. \edit{In retrospect, the topical diversity in the first study, both between different sources and within tweets published by a single source, does not necessarily portray systematic biases which would reduce the perceived credibility of a source. In the next study, we explore this issue by fixing the textual content shown to users and manipulating only the emotional valence of facial expressions of images.}

\begin{figure}[t]
  \centering
    \includegraphics[width=0.6\columnwidth]{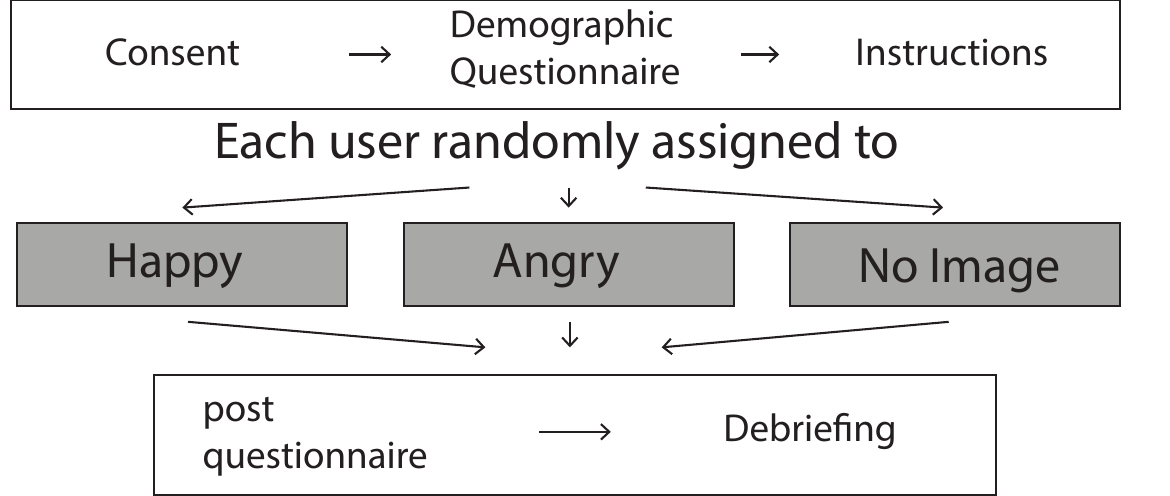}
  \caption{Study 2 conditions and process}
  \label{fig:rq2-design}
  \vspace{-5pt}
\end{figure}


\section{Study 2}
In study 1, the content shown to users was not limited to specific topics and we found that users' judgments were likely influenced by different topical focuses of sources. To better evaluate the impact of emotional facial expressions on users' judgments, we shifted the focus of study 2 to explore the effects of \edit{mainstream} sources' systematic emotional portrayal of specific politicians. To address such a scenario, in all experimental conditions of this study 2, users received the same 8 sets of tweet texts about 8 different politicians. Per random assignment, users received either angry or happy images of the politicians. We included no images as a control condition in which users received the same sets of tweets without any accompanying images of those politicians.

\begin{figure}[t]
  \centering
    \includegraphics[width=0.8\columnwidth]{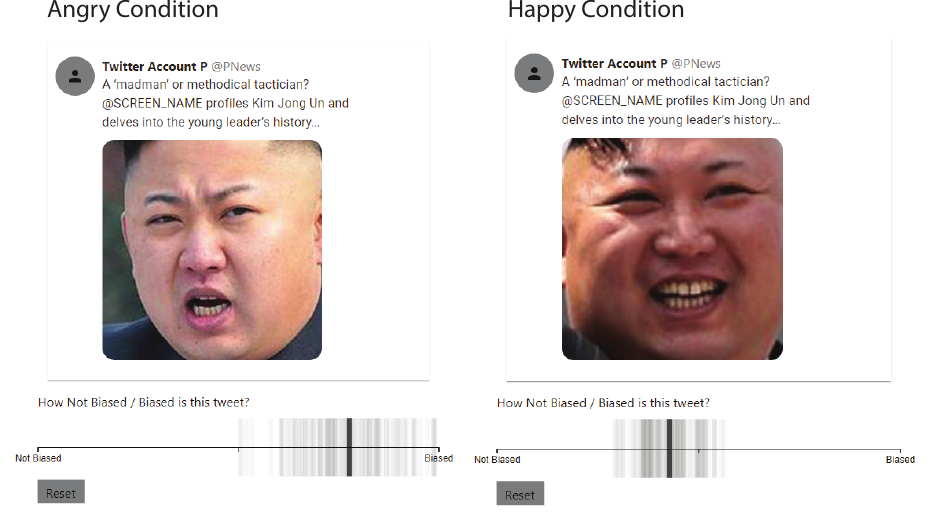}
  \caption{Sample of study 2 (Angry and Happy) conditions for one tweet related to Kim Jong Un. }
  \label{fig:rq2-sample}
  \vspace{-10pt}
\end{figure}

\begin{table}[t]
\scriptsize	
\sffamily
\centering
\caption{Politicians selected for study 2}
\label{tab:study2Identities}
\begin{tabular}{|l|l|}
\hline
\rowcolor[HTML]{656565} 
{\color[HTML]{FFFFFF} Politician}       & {\color[HTML]{FFFFFF} Notes}                 \\ \hline
\rowcolor[HTML]{FFFFFF} 
{\color[HTML]{000000} Donald Trump}    & {\color[HTML]{000000} Former President of the US}     \\ \hline
\rowcolor[HTML]{C0C0C0} 
{\color[HTML]{000000} Hillary Clinton} & {\color[HTML]{000000} Former US Secretary of State}    \\ \hline
\rowcolor[HTML]{FFFFFF} 
{\color[HTML]{000000} Barack Obama}    & {\color[HTML]{000000} Former president of the US}      \\ \hline
\rowcolor[HTML]{C0C0C0} 
{\color[HTML]{000000} Theresa May}     & {\color[HTML]{000000} Former prime minister of the UK} \\ \hline
\rowcolor[HTML]{FFFFFF} 
{\color[HTML]{000000} Emanuel Macron} & {\color[HTML]{000000} President of France}   \\ \hline
\rowcolor[HTML]{C0C0C0} 
{\color[HTML]{000000} Angela Merkel}  & {\color[HTML]{000000} Former Chancellor of Germany} \\ \hline
\rowcolor[HTML]{FFFFFF} 
{\color[HTML]{000000} Kim Jong Un}     & {\color[HTML]{000000} Supreme Leader of North Korea}   \\ \hline
\rowcolor[HTML]{C0C0C0} 
{\color[HTML]{000000} Vladimir Putin} & {\color[HTML]{000000} President of Russia}   \\ \hline
\end{tabular}
\end{table}

 To curate the stimuli for the present study, the tweets dataset from the previous study was filtered to tweets that included mentions of 8 different prominent politicians (see Table \ref{tab:study2Identities}). Users were instructed that each set of tweets mentioning each politician were from a single source. To limit the impact of emotional text content on users’ judgments, we downselected tweets with the following steps: we first conducted sentiment analysis on the tweet texts using Vader Sentiment \cite{hutto2014vader}. Next, for each set of tweets, we selected tweets with the highest neutral sentiment scores. Finally, we qualitatively evaluated and removed tweets with inaccurate scores from the sentiment analysis library. This resulted in 30 to 40 tweets mentioning each politician from mainstream news sources that were mostly of neutral tone.
 The images for this study were manipulated in a between-subjects manner such that users saw either happy, angry, or no images of each politician. For example, a user in the Happy condition, viewed tweets mentioning Kim Jung Un accompanied by happy images of him, while a user in the Angry condition evaluated the same tweets but with angry images of him, and the control (no-image) condition viewed no images (see Figure \ref{fig:rq2-sample} for one of the example tweets). 

The procedures of the study were equivalent to Study 1, with one exception: To measure the interaction between users' prior attitude towards each politician and the experimental conditions, for each set of tweets users first answered two questions in a pop-up form about their familiarity and favorability towards that politician on 5-point Likert scales.

\subsection{Hypotheses}
Since study 2 is a continuation of study 1, we expected to see a similar effect of angry emotions on content bias. We also hypothesized that users in the angry condition would rate sources as less credible. Guided by research on motivated reasoning \cite{kahan2012ideology}, we expected to observe an interaction between users’ prior favorability of different politicians and sources' negative and positive visual bias towards those politicians. More specifically, we hypothesized that favorability negatively interacts with angry emotion and positively interacts with happy emotion to predict users perceived bias and credibility. Furthermore, we also investigate the impact of users’ familiarity with each politician on their judgments. 

\subsection{Dependent \& Independent Variables} The dependent variables in study 2 are identical to study 1. We considered four total dependent variables(DV): (1) the tweet bias choice (bounded value between [0,1]), (2) uncertainty range around tweet bias (bounded value between [0,1]), (3) source credibility choice (bounded value between [0,1]), (4) uncertainty around source credibility choice (bounded value between [0,1]). For our independent variables (IV), we included the image emotion condition (angry, happy, or no image), as well as users’ prior favorability and familiarity towards each politician. 

\subsection{Model specification}
For each model, we included users’ unique IDs and the politician's name as random effects. After comparing multiple model specifications using AIC, we also included interaction terms between users’ favorability and familiarity of each politician with the image emotion. The omitted reference conditions are image emotion = no image.

\subsection{Participants}
\begin{figure}[t]
  \centering
    \includegraphics[width=0.9\columnwidth]{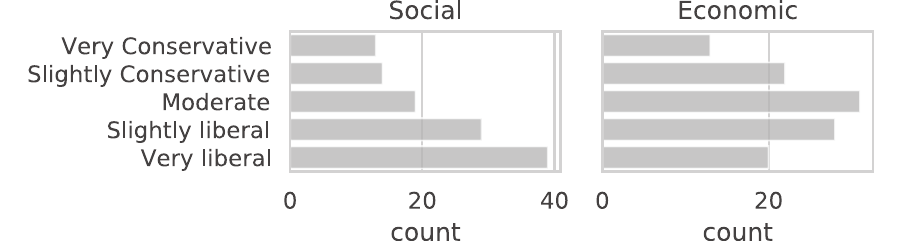}
  \caption{Participants' responses to questions "How would you describe your political outlook with regard to economic / social issues?" for Study 2.
  }
  \label{fig:rq2-political}
\end{figure}

In study 2, we recruited a total of 126 participants. The average age of participants was 35 years old. 81 participants were recruited from Amazon Mechanical Turk and received a 2 dollar incentive. To get closer to our pre-registration target, the rest of the participants were university students who received either research or course credit for their participation. Per our pre-registration, we excluded responses from 12 participants with missing responses (due to unexpected technical difficulties) resulting in 114 accepted responses (54 woman, 59 man, and 1 prefer not to say; 87 white, 10 African American, 7 east Asian, 5 Hispanic, 3 Other Asian and 2 middle eastern). \edit{Participants were highly liberal in regards to social political issues and more balanced in regards to economic issues (See Fig \ref{fig:rq2-political})}. 34 participants were randomly assigned to the angry condition, 42 participants were assigned to the happy condition, and the remaining 38 were assigned to the no image condition. Participants took an average of 21 minutes to complete the study.

\subsection{Results}

\subsubsection{Content bias:}
\begin{figure}[t]
  \centering
    \includegraphics[width=0.9\columnwidth]{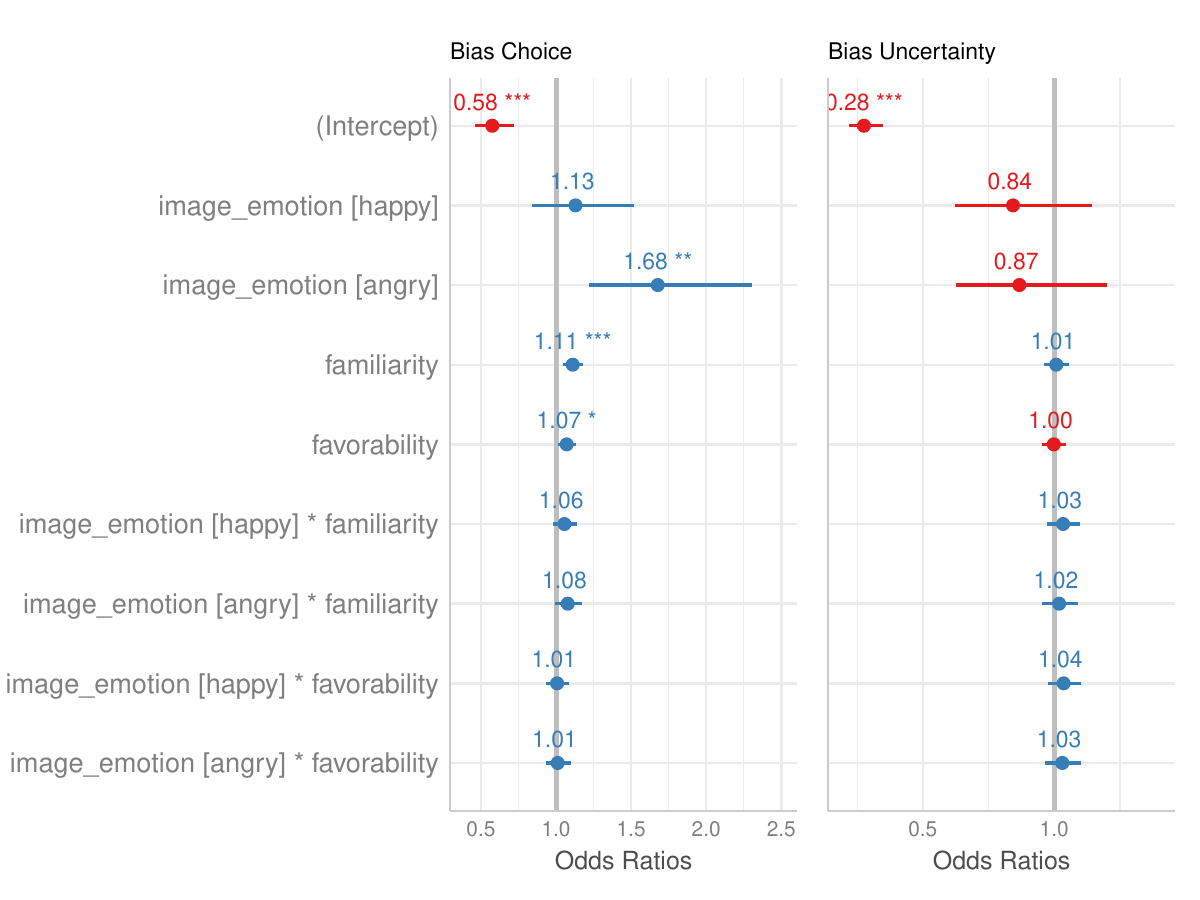}
  \caption{
Study 2 fixed effects Odds Ratios for bias choice (left) and bias uncertainty (right). Error bars indicate 95\% confidence intervals. 
Asterisks indicate statistical significance than zero using p-values: *** 99.9\%, ** 99\%, * 95\%. 
  }
  \label{fig:rq2-bias}
\end{figure}

We used two mixed-effects beta regressions to study the effects of experimental conditions on users’ judgment and uncertainty of bias of individual tweets (see fig \ref{fig:rq2-bias}-left). We found that users viewing tweets with angry images rated tweets as more biased in comparison to when no images are shown to users ($\beta = 0.511 \; [0.18, 0.84], z =  3.031, p < .01$). This is in line with our hypothesis that when the content of messages is the same, angry imagery leads to an increase in perceived bias. Similar to study 1, we did not observe any effect in the happy condition. Moreover, we did not find any evidence for an interaction between favorability/familiarity and image emotion. We did however, observe an overall positive effect of familiarity on users' perceived bias ($\beta = 0.115 \; [0.05,  0.17], z =  3.82, p < .001$). We also observed a small positive effect of favorability on perceived bias ($\beta = 0.074 \; [0.01, 0.13], z =  2.471, p < .05$). We did not observe any significant effects of any of the independent variables on users’ uncertainty (See Fig \ref{fig:rq2-bias}).

\begin{figure}[t]
  \centering
    \includegraphics[width=0.9\columnwidth]{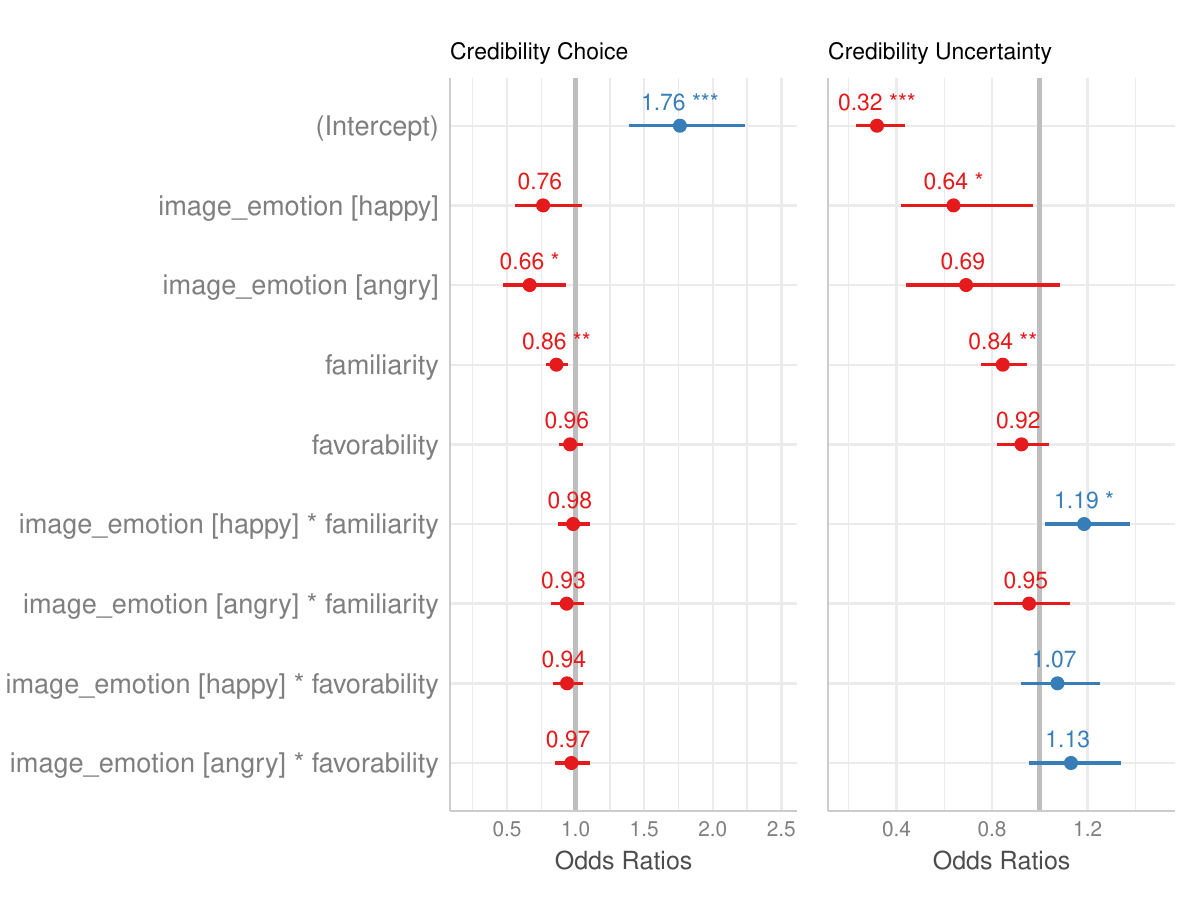}
  \caption{
Study 2 fixed effects Odds Ratios for credibility choice (left) and credibility uncertainty (right). Error bars indicate 95\% confidence intervals. 
Asterisks indicate statistical significance than zero using p-values: *** 99.9\%, ** 99\%, * 95\%. 
  }
  \label{fig:rq2-credibility}
\end{figure}

\subsubsection{Source credibility:} We used mixed-effects beta regression to investigate the effects of study conditions on users’ judgments about source credibility (see Fig \ref{fig:rq2-credibility}-left). We found that users rated sources as less credible when tweets were accompanied with angry facial expressions ($\beta = -0.397 \; [-0.73, -0.05], z =  -2.308, p < .05$). This finding is in line with our hypothesis that angry facial expressions will lead to a decrease in the perceived credibility of sources. We also find that familiarity also decreases  users' perceived credibility of sources ($\beta = -0.153 \; [-0.245, -0.06], z =  -3.256, p < .01$). Again, here we did not observe an interaction effect between favorability and the image emotion.

For the mixed effect model of users’ uncertainty, we observed a significant effect of happy emotion on users’ judgment uncertainty ($\beta = -0.44 \; [-0.8 -0.02], z =  -2.09, p < .01$). We also observed that in cases that users' familiarity with subjects was higher, users had more certain judgments ($\beta = -0.168 \; [-0.28 -0.05], z =  -2.9, p < .01$). We also observed a positive interaction between happy emotion and users' familiarity ($\beta = 0.16 \; [0.024  0.31], z = 2.224, p < .05$), such that there was greater uncertainty in the happy condition for more familiar subjects.

\subsection{Qualitative analysis of users’ comments} In study 2, we collected a total of 881 comments from 116 users. Similar to study 1, we analyzed users’ descriptions of the rationale behind their judgments using NMF topic modeling \cite{cichocki2009fast} and thematic analysis of the documents most representative of each topic. This helped us categorize the 20 extracted topics into 5 higher-order themes. Four of the themes were similar to the ones we extracted from study 1 and discussed cues found in the text of the tweets. Here again, we will offer a qualitative overview of users' comments about images.

We identified three topics that contain descriptions from users related to visual information. Topic 19 with 71 comments from 34 users and Topic 14 with 47 comments from 24 users include mentions of facial expressions, angry emotions, and unflattering portrayals. For example, a user found the text of tweets mostly unbiased, and explained how portrayed facial expressions of Hillary Clinton was the basis for their judgment: \textit{``While the text didn’t involve much biased words, the usage of certain images of Hillary Clinton depicting her facial expressions in an array of negative emotions showed a biased view
.''}


Some users also found a combination of text with “weirdly close up” and “unflattering” images leading them to believe a source is less credible: \textit{“The texts were mostly bland, except for “...what do you think”, which is a tabloid-like phrase for me. Photos were weirdly close-up facial views that were generally unflattering, which makes me wonder a bit about credibility...”}. Another interesting set of comments were about users perceiving the images as not correlating with the tweets: \textit{“Most of the tweets were very normal, but there were a couple that had angry Macron pictures that did not correlate with the headlines presented.”} And \textit{“The majority of tweets were unbiased with their headlines.  Some of the pictures might have been a bit questionable.”}

A group of comments described how images did not influence users' judgments. Topic 16 with 66 comments from 32 users, includes examples of such commentary. For example, for tweets about Kim-Jung Un, a user mentioned: 
\textit{“The images were not the best images nor were they the worst images of him, but there was still a negative bias.”} Another user mentioned how images and tweets related to Donald Trump were both neutral, and honest looking: \textit{“The tweets seemed to be honest and state the honest news about what is happening while the images associated with them.”}

\begin{figure}[t]
  \centering
    \includegraphics[width=1.0\columnwidth]{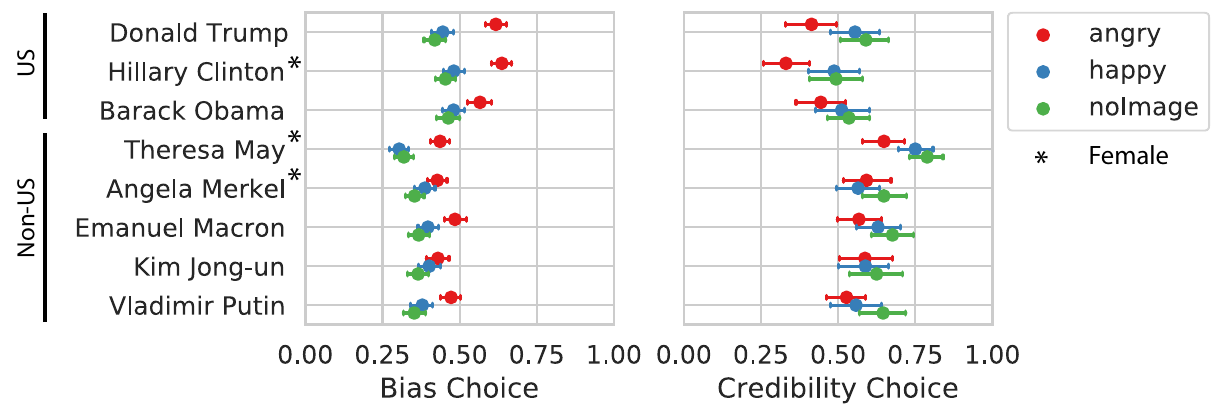}
  \caption{
  Mean and 95\% bootstrapped confidence interval for bias and credibility choice for each set of tweet about each politician.
  }
  \label{fig:rq2-pointplot}
\end{figure}

\subsection{Study 2 discussion}


We asked participants to judge sources that each focused on a specific politician. The text was constant for all users, while we manipulated the tweets to include either negative (angry) or positive (happy) facial expressions of those politicians or to include no images as a control condition. We found that in comparison to users in the no image condition, users in the angry condition found the content to be more biased and the sources to be less credible. We did not find a reverse effect of happy images on users’ judgments. The difference between the happy and angry condition in our study could be better explained by a study on perceptions of negative or positive portrayals of politicians, in which Lubinger and Brantner found that participants mostly agreed on what constituted as negative, but perceptions of positive portrayals varied widely \cite{lobinger2015likable}. This suggests that negative portrayals might be more salient and more likely to be commonly agreed upon and thus might have a stronger effect on users’ judgments. Additionally, users’ comments also included many mentions of angry, negative, or unflattering portrayals for these politicians, while we did not observe mentions of positive bias in users' comments. 

\begin{figure}[t]
  \centering
    \includegraphics[width=0.9\columnwidth]{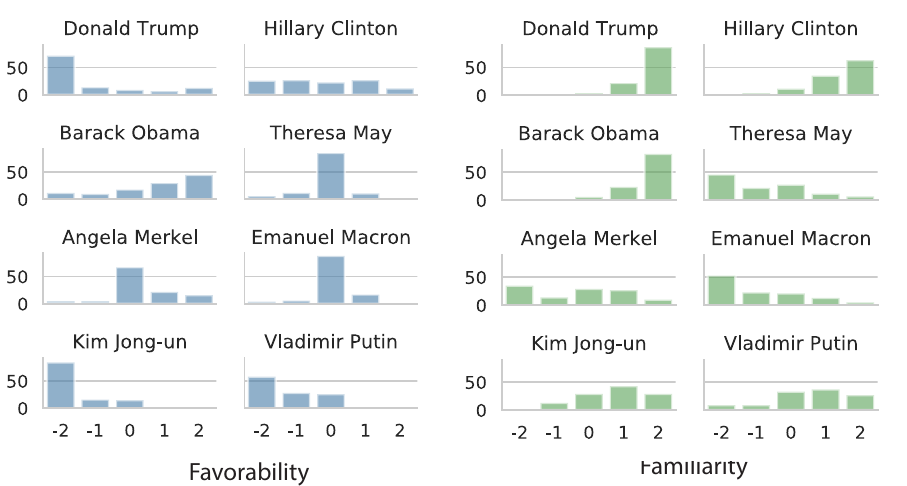}
  \caption{
  Count of Favorability and Familiarity responses (Study 2).
  }
  \label{fig:rq2-favefam}
  \vspace{-10pt}
\end{figure}

Although the effect is rather small, we observed that users rate tweets about politicians they are more familiar with as more biased and to come from less credible sources \edit{(See Fig \ref{fig:rq2-favefam} for counts of favorability and familiarity responses for each politician)}. Recent work on judgments on misinformation suggests that prior exposure and familiarity with misinformation increases the perceived accuracy of content in which users would have to rely on their memory to assess the accuracy of news headlines or articles \cite{pennycook2018prior}. Since to assess the credibility and bias of anonymous content, users would not rely on their memories, we suspect that the impact of familiarity on these judgments might be different. Our comment analysis highlighted that users often rely on more analytical approaches and different  heuristics such as negativity, word usage, or emotions in facial expressions to judge bias and credibility of sources. An explanation for the effect of familiarity on users’ judgment about bias and credibility could be that when users are more familiar with politicians, they might be more sensitive to the details of texts and images of content they view and therefore possibly more likely to identify cues impacting their credibility and bias judgments. 

Finally, we observed a small overall effect of favorability on users' perception of tweet bias, but we did not observe an interaction between favorability and emotions in images. Although we elicited users' self-described political orientation, we elected to instead use favorability and familiarity of specific politicians to more directly study the impact of users' existing attitudes on their credibility and bias judgments. We expected that if a user is engaged in motivated reasoning, they would find attitude-consistent portrayals (e.g., an angry portrayal of a politician they dislike) as a cue for credibility, while portrayals that conflict with preexisting attitudes would be associated with lower credibility judgments. However, our results do not provide evidence that, in such experimental settings, edit{emotions in facial expressions impacts} users who might be engaged in motivated reasoning based on their favorability towards politicians. This result is in line with work showing that motivated reasoning is not a primary factor in users' judgments when they are asked to evaluate news content \cite{pennycook2019lazy}. 

\section{General Discussion and Conclusion}

 Study 1 showed that angry facial emotions lead to higher ratings of content bias, but not lower credibility ratings for sources. However, these judgments were possibly impacted by differences in the textual content and topics across the sources we used. In  study 2 we controlled for the \edit{textual content} while investigating the systematic negative or positive visual portrayals of politicians on users perceptions of credibility and bias. Our results provide partial evidence for our hypothesis: systematic negative visual treatments of politicians led to both higher perceived content bias and lower perceived source credibility. However, we did not find strong evidence for our hypotheses around the interactions between users' prior favorability of politicians and sources' systematic negative or positive visual portrayal of those politicians.
 
We also explored the impact of different study conditions on users' certainty around their judgments. 
In study 1, misinformation sources were associated with more certain credibility judgments. One possible interpretation for this effect is that many misinformation sources are more likely to use content with more suspicious language and images \cite{volkova2019explaining, volkova2017separating}. In study 2, we observed that users were more certain in their credibility judgments when they were more familiar with a politician.
Users with greater direct knowledge of a politician may have been more sensitive to the accuracy of the content and were therefore able to make more confident credibility judgments.
 
Although these results provide evidence of the impact of emotional facial expressions in images on users' judgments, we found that users also rely heavily on other cues in the tweet content such as political orientation and opinionated language. It is also important to note that there are many more aspects related to emotions in images that remain to be studied. A comparison between angry, happy, and neutral emotions is an appropriate comparison which we omitted due to the difficulty of curating a balanced dataset with neutral images of all eight politicians. Moreover, within each of the angry or happy emotional categories, there are finer distinctions between different emotional facial expressions, ranging from extremely angry/happy to subtle frown/grin that might impact users' judgments. Comments about satirical images point to the importance of how subtle changes in facial expressions could impact users' perceptions of emotional facial expressions. There are also other emotion dimensions such as sadness, surprise, fear, or disgust that might potentially impact users' judgments. Many of these subtle differences might be present in our  stimuli and may have impacted participants' judgments. Finally, facial expressions rarely contain one unique emotion and subtle changes might communicate different meanings to individuals. \edit{In addition to facial expressions, prior research has established the influence of other visual features (including body expression and scene information) \cite{kret2013.00810, Kret2012a, Kret2013} on the perception of emotions. Although these factors were outside of scope of the two studies reported in this paper, they are promising next steps to investigate.} \edit{In future studies,} Generative Adversarial Neural (GAN) networks \cite{goodfellow2020generative} can be used to produce image datasets with finer control over the facial expressions or other emotional content \cite{todorov2021data}. 

We acknowledge some general limitations in the design and execution of our studies. \edit{To remove potential confounds introduced by body posture and image backgrounds, we cropped all images to include only faces. However, this decision might have a general negative effect on users' bias judgments as seen in some of participants' comments and prior research identifying face size as a specific type of visual bias \cite{peng2018same}. Additionally, for users' judgments we did not measure differences between positive and negative content bias. Eliciting valenced judgments about content bias may provide more insight into how these judgments depend on users' pre-existing attitudes or familiarity with a subject. In the future it is important to control for and study impact of factors such as face size and positive/negative bias on users' judgments.} 
Moreover, To limit the impact of users’ preconceived notions of sources, we masked all account names from our stimuli. Even though encountering new and unknown sources is a realistic scenario and especially important in the context of misinformation, in many cases users might have a self-selected set of sources they trust and refer to. It is possible that users' prior knowledge of sources significantly impacts how they evaluate content. For example, a user who trusts Fox News or Washington Post might be less sensitive to any visual bias portrayed by these sources towards a politician. An important future step for our research is to investigate how users update their trust in familiar sources when they observe a systematic usage of biased visual content. \edit{Other factors such as gender and political orientation of politicians are also likely to impact users' judgments of bias and credibility and require closer attention in future studies. Although we did not explicitly pre-register and control for such factors in our experiments, we did not observe noticeable differences between politicians in our stimuli (See Fig \ref{fig:rq2-pointplot}).}  

\section{Broader perspective and ethics}
This study was conducted in experimental settings. The procedures of the study were approved by our institutions IRB, every participant signed an informed consent, and throughout the study the participants identities remained anonymous. \edit{We believe our results could potentially be used by malicious actors to further influence consumers' judgments. However, as most social media interactions are driven by polarized outrage \cite{rathje2021out},  these results can engage the community in ethical discourse on curation of sensationalized visual media by mainstream and misinformation organizations}. Moreover, to learn about the true impacts of content on users' beliefs and attitudes, we believe that results of experimental settings should be repeated ``in the wild'' \cite{mosleh2021field}. Such in-the-wild experiments, however, require attention to both platform rules (in our case, Twitter), and ethical guidelines of interacting with users on social media. These ethical questions for social media experiments are critical, as it is difficult to maintain anonymity on these platforms. In the future, as we move towards combating harmful information on social media, such ethical considerations become truly essential.  

\section{Conclusion}

Across two consecutive preregistered studies with a total of 207 participants, we find evidence that negative (angry) facial expressions in social media news images lead to an increase in users’ perception of bias in content. We also found that users judge sources to be less credible when they observe a potentially systematic negative portrayal of different politicians. While this paper was motivated by the prevalence and impact of misinformation on our democracies and societies, our current globally politicized and polarized political ecosystem calls for a more critical view on the whole media landscape. This paper highlights the importance of moving beyond accuracy of content in the constant \edit{struggle to build trust in credible news sources}. Sensationalized news content, even in subtle ways such as emotional facial expressions, might impair users' trust in otherwise trustworthy news. These effects can lead to harmful outcomes evident in many current issues such as vaccine hesitancy and polarization. We would like to take this opportunity to call the research community to expand their focus on the broader impacts of news contents on users beliefs and attitudes. 

\bibliography{references.bib}
\end{document}